\documentclass[aps,prb,reprint,showpacs,amsmath,amssymb,floatfix,footinbib,citeautoscript,superscriptaddress]{revtex4-1}
\usepackage{graphicx}
\usepackage{color}
\usepackage{amstext}
\usepackage{amsmath}
\usepackage{amssymb}
\usepackage{bm}
\usepackage{graphics}
\usepackage{dsfont}
\usepackage{subfigure}
\usepackage{ulem}

\newcommand{\Ref}[1]{Ref.~\onlinecite{#1}}  
\newcommand{\Sec}[1]{Sec.~\ref{#1}}  
\newcommand{\Equ}[1]{Eq.~(\ref{#1})}  
\newcommand{\Fig}[1]{Fig.~\ref{#1}}  

\def\Tr{\mathop{\mathrm{Tr}}}
\newcommand\E{\mathrm{e}}
\newcommand\I{\mathrm{i}}

\newcommand{\pd}{\phantom{\dagger}}
\newcommand{\dd}{\mbox{d}}

\emergencystretch=\maxdimen
\begin{document}

\title{Spontaneous particle-hole symmetry breaking\\of correlated fermions on the Lieb lattice}
\author{Martin Bercx}
\affiliation{Institut f\"ur Theoretische Physik und Astrophysik, Universit\"at W\"urzburg, 97074 W\"urzburg, Germany}
\author{Johannes S. Hofmann}
\affiliation{Institut f\"ur Theoretische Physik und Astrophysik, Universit\"at W\"urzburg, 97074 W\"urzburg, Germany}
\author{Fakher F. Assaad}
\affiliation{Institut f\"ur Theoretische Physik und Astrophysik, Universit\"at W\"urzburg, 97074 W\"urzburg, Germany}
\author{Thomas C. Lang}
\affiliation{Institut f\"ur Theoretische Physik und Astrophysik, Universit\"at W\"urzburg, 97074 W\"urzburg, Germany}
\affiliation{Institute for Theoretical Physics, University of Innsbruck, 6020 Innsbruck, Austria}

\begin{abstract}
We study spinless fermions with nearest-neighbor repulsive interactions ($t$-$V$ model) on the two-dimensional three-band Lieb lattice. At half-filling, the free electronic band structure consists of a flat band at zero energy and a single cone with linear dispersion. The flat band is expected to be unstable upon inclusion of electronic correlations, and a  natural channel is charge order. However, due to the three-orbital unit cell, commensurate charge order implies an imbalance of electron and hole densities and therefore doping away from half-filling. Our numerical results show that below a finite-temperature Ising transition a charge density wave with one electron and two holes per unit cell and its partner under particle-hole transformation are spontaneously generated. Our calculations are based on recent advances in auxiliary-field and continuous-time quantum Monte Carlo simulations that allow sign-free simulations of spinless fermions at half-filling. It is argued that particle-hole symmetry breaking provides a route   to access levels of finite doping, without introducing a sign problem.
\end{abstract}

\pacs{71.10.-w,71.10.Hf,02.70.Ss}

\maketitle

\section{Introduction} \label{sec:intro}

Controlled and approximation free quantum Monte (QMC) simulations have been long considered to be limited to certain parameter regimes -- unless one is willing to abandon the advantage of polynomial scaling of the algorithm and suffer the notorious fermionic sign problem. The origin and manifestation of the sign problem varies in different QMC algorithms. In the auxiliary-field approach \cite{Blankenbecler81} the absence of sign problem relates to symmetry properties of the action in complex fermion \cite{Wu04}, or Majorana \cite{Li15,Li16,Wei16} representations. A prominent niche are  particle-hole symmetric systems (half-filling) with repulsive interactions which do not suffer from the sign problem for non-frustrated hopping and interactions \cite{Hohenadler14}. In this manuscript, we show that insight into properties away from half-filling may be obtained from finite-size simulations at particle-hole symmetry. The fact that particle-hole symmetry may only be broken spontaneously in the thermodynamic limit allows for QMC simulations, which sample the phase space of finite doping while remaining sign problem free.

We investigate the electronic correlation effects on the face-centered square, or Lieb lattice. The three-orbital  unit cell structure of the lattice enables perfect destructive interference of electronic hopping processes, which generates localized states at zero energy that form a dispersionless band. 
While the Lieb lattice is reminiscent of the CuO$_2$ plane of high temperature superconductors, here we want to investigate the fundamental problem of the electronic instabilities when two linearly dispersing and one flat band meet at the Fermi level at a singular point in the Brioullin zone. An odd number of orbitals per unit cell make the system inherently prone to order, such as ferromagnetism in the case of SU(2)-symmetric electrons \cite{Lieb89}.
 In case of interacting spinless electrons, the system either has an unique or a doubly degenerate ground state. \cite{Wei15}
 The robustness of the flat band of the Lieb lattice has been investigated for magnetic fields \cite{Goldman11}, spin-orbit interactions \cite{Goldman11, Weeks10,Beugeling12}, local \cite{Scalettar91,Nie16,Noda15,Costa16} and inter-site \cite{Jaworowski15,Dauphin16} Coulomb repulsion, attractive interactions \cite{Iglovikov14,Julku16}, as well as disorder \cite{Nita13}. 
Topological surface states also exhibit flat bands that are susceptible to interactions \cite{Hofmann2016,Honerkamp2000,Potter2014,KopninVolovikPRB11,graphene_edge_magnetism,Feldner2010,Feldner2011,tangFu_natPhys14,Roy2014,Li2013}. 
\begin{figure}[tp]
	\includegraphics[width=\columnwidth]{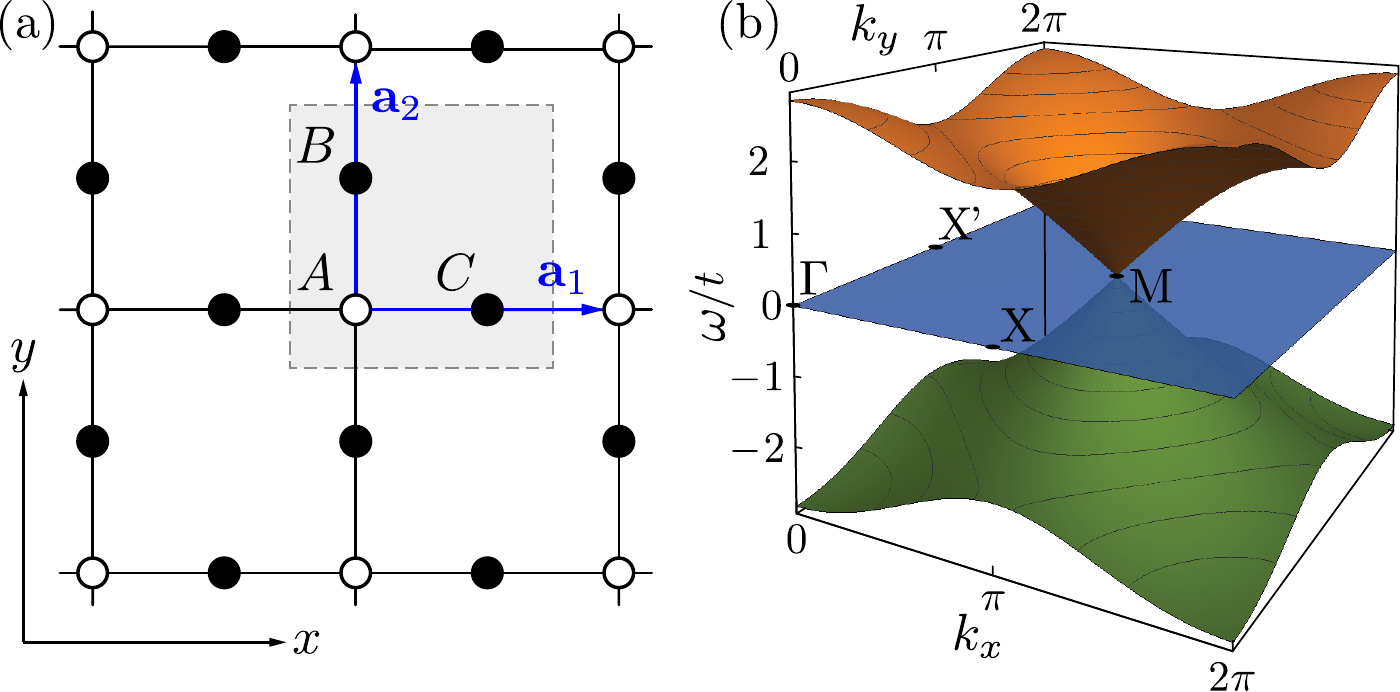}
	\caption{The bipartite lattice structure of the Lieb lattice (a) with the three orbitals $A$, $B$ and $C$ per unit cell indicated in gray and the dispersion of the non-interacting system (b) which features a single linearly dispersing cone at the corner of the Brillouin zone.
	\label{Fig:FreeSystem}}
\end{figure}
A recent review on strongly correlated flat-band systems is presented in Ref.~\onlinecite{Derzhko15}. The Lieb lattice geometry has recently been realized in optical lattices \cite{Mukherjee15,Vicencio15}. Populating these lattices with spin-polarized (spinless) fermionic atoms would allow to directly investigate the scenario presented in this manuscript.

Here, we apply continuous-time QMC simulations, auxiliary-field QMC simulations and exact diagonalization to investigate the correlation effects of spinless electrons subject to nearest-neighbor Coulomb repulsion on the Lieb lattice. Both QMC methods conserve the particle number and allow for simulations free of the sign problem at the particle-hole symmetric point. As argued in  Refs. \onlinecite{Feldbach03,Assaad16}, particle-hole symmetry corresponds to an Ising order parameter, which in two dimensions can spontaneously condense at finite temperature. In the aforementioned references particle-hole symmetry breaking amounts to specific charge ordering below the transition, but the system remains half-filled. Here, the situation is notably different due to the fractional number of spinless fermions ($3/2$) per unit cell. In fact the  symmetry broken states correspond to charge ordered states at filling factors $1/3$ and $2/3$ which are connected by a particle-hole transformation. During the stochastic sampling process in a particle-hole symmetric QMC simulation, both realizations of the broken symmetry are equally sampled. Nevertheless, the spontaneous symmetry breaking can be inferred from the finite-size extrapolation of correlation functions.

The main results and structure of the  manuscript are the following. In Sec.~\ref{sec:model}, we introduce the $t$-$V$ Hamiltonian, briefly review the non-interacting model and discuss the symmetries, which enable sign problem free QMC simulations. In Sec.~\ref{sec:method}, we then introduce continuous-time QMC methods as well as the auxiliary-field QMC algorithm for spinless electrons. For this specific flat band model, the choice of the algorithm turns out to be crucial since depending upon the formulation heavy-tailed distributions for certain observables occur. Section~\ref{sec:results} contains exact diagonalization and QMC results. From real-space charge-charge correlations we extract the emergence of commensurate charge order below the critical temperature. We study the critical behavior by performing a finite-size scaling analysis of the order parameter and show that the phase transition belongs to the two-dimensional Ising universality class. We also discuss the single-particle spectral function. In Sec.~\ref{sec:conclusion} we conclude and discuss possible implications of our result, in particular the possibility of using spontaneous particle-hole symmetry breaking to access finite doping without introducing a sign problem.

\section{Model \& Symmetries} \label{sec:model}

We study spinless fermions on a two-dimensional Lieb lattice [cf. \Fig{Fig:FreeSystem}(a)] interacting via a nearest-neighbor Coulomb repulsion described by ${\mathcal{H}_0 + \mathcal{H}_V}$, with
\begin{eqnarray}
  \label{Eq:HFree}
  \mathcal{H}_0 & = & -t\sum_{\langle {\bf i},{\bf j} \rangle} (c^\dag_{\bf i} c_{\bf j}^{\phantom{\dag}} +\text{H.c.})\;,\\
  \mathcal{H}_V & = &  V\sum_{\langle {\bf i},{\bf j} \rangle} \left(n_{\bf i}^{\phantom{\dag}}-\frac{1}{2}\right)\left(n_{\bf j}^{\phantom{\dag}}-\frac{1}{2}\right) \label{Eq:HV}\\
   & = & - \frac{V}{2} \sum_{\langle {\bf i},{\bf j} \rangle} \left[\left(c_{\bf i}^{\dagger} c_{\bf j}^{\phantom\dagger} + \text{H.c.}\right)^2 - \frac{1}{2}\right]\;,\label{Eq:HV2}
\end{eqnarray}
where $c^\dag_{\bf i}$ creates a spinless electron on lattice site ${\bf i}$, $t$ denotes the hopping amplitude and $V$ the interaction strength. The Fourier transformation of the non-interacting part of the Hamiltonian to momentum space generates $\mathcal{H}_0=\sum_{\bf{k}}\Psi_{\bf{k}}^\dagger \, \bf{H}({\bf{k}}) \, \Psi_{\bf{k}}^{\phantom{\dagger}}$ with ${\Psi_{\bf{k}}^\dagger=\left( c^{\dagger}_{A{\bf k}},c^{\dagger}_{B{\bf k}},c^{\dagger}_{C{\bf k}}\right)}$ and expanding it to leading order around the $\bf{M}$-point at $(\pi,\pi)$ gives
\begin{equation}
   \label{eqn:ham_matrix_linear}
   {\bf H}({\bf M} + {\bf q}) = -t ( q_{x} \, \mathbf{S}_{x} + q_{y} \, \mathbf{S}_{y}) + \mathcal{O}(q^{2})\;.
\end{equation}
Here the $\mathbf{S}$ matrices are the spin $S=1$ representation of the SU(2) Lie-Algebra \cite{Green10,Goldman11,Tsai15}. The eigenvalues are given by $\{0,\pm \left| \bf q \right| \}$ such that the Hamiltonian hosts a zero energy flat band and two linear dispersing modes as it is typical for a spin-$1$-cone \cite{Green10}. The spectrum for the whole Brillouin zone is depicted in \Fig{Fig:FreeSystem}(b).

The zero energy mode is not a coincidence, but rather a consequence of the particle-hole symmetry in \Equ{Eq:HFree} with the corresponding transformation
\begin{equation}
   c^\dag_{\bf i}\rightarrow \left\{
   \begin{array}{cl}
      -c_{\bf i}^{\phantom{\dag}}, & \quad {\bf i}\in\mbox{sublattice $A$}\\
      \phantom{-}c_{\bf i}^{\phantom{\dag}}, & \quad {\bf i}\in\mbox{sublattices $B$, $C$}.
   \end{array}
   \right.
   \label{Eq:PHT}
\end{equation}
It guarantees that every eigenvalue comes as a $\pm \omega$ pair, hence there has to be a zero energy mode for an odd number of degrees of freedom within one unit cell.

The system's particle-hole symmetry combined with the global U(1) symmetry (charge conservation) generates a O(2) symmetry in a suitably chosen Majorana basis, as we will show below: Recasting the interaction term [\Equ{Eq:HV}] as a square of the hopping term [\Equ{Eq:HV2}] allows us to perform a Hubbard-Stratonovich transformation, which explicitly makes use of the O(2) symmetry and thereby enables a sign problem free formulation of the auxiliary-field QMC algorithm presented in \Sec{sec:afqmc}. To show the invariance of the Hamiltonian, we first diagonalize the unitary part of the above particle-hole transformation by introducing the operators ${d_{\bf i}=\I c_{\bf i}}$ on sublattice $A$ and ${d_{\bf j}=c_{\bf j}}$ on sublattices $B,C$. Secondly we define the Majorana operators $\gamma_{\bf i}$ and $\eta_{\bf i}$ as follows 
\begin{equation}
   \label{Eqn:majorana}
   d^{\phantom\dagger}_{\bf i} = (\gamma_{\bf i} +\I\eta_{\bf i})/2\;,\quad 
   d^{\dagger}_{\bf i} = (\gamma_{\bf i} - \I\eta_{\bf i})/2\;.
\end{equation}
The fermion commutation relation, ${\{c_{\bf i},c_{\bf j}^{\dagger}\}=\delta_{\bf ij}}$, fixes the Majorana commutation relations to ${\{\gamma_{\bf i},\gamma_{\bf j}\}=\{\eta_{\bf i},\eta_{\bf j}\}=2\delta_{\bf ij}}$ and ${\{\gamma_{\bf i},\eta_{\bf j}\}=0}$. The bond density can then be expressed in terms of the Majorana fermions
\begin{equation}
   \label{Eqn:Hmajorana}
   c_{\bf i}^{\dagger} c_{\bf j}^{\phantom\dagger} + c_{\bf j}^{\dagger} c_{\bf i}^{\phantom\dagger}
   = \I (d_{\bf i}^{\dagger} d_{\bf j}^{\phantom\dagger} - d_{\bf j}^{\dagger} d_{\bf i}^{\phantom\dagger})
   = \frac{\I}{2}( \gamma_{\bf i}\gamma_{\bf j} + \eta_{\bf i}\eta_{\bf j})\;,
\end{equation}
which is invariant under O(2) transformations of the Majorana basis.

The rotation group O(2) consists of rotations and reflections represented by the rank two matrices $\mathbf{R}(\theta)$ and $\mathbf{P}$, respectively. The bond densities in the Majorana representation in Eq.~(\ref{Eqn:Hmajorana}) and consequently the Hamiltonians of Eq.~(\ref{Eq:HFree})  and (\ref{Eq:HV2}) are invariant under the O(2) transformations $\mathbf{R}(\theta) \otimes \mathds{1}$ and $\mathbf{P}\otimes \mathds{1}$ acting globally on all sites. The group of special rotations SO(2) constitutes a subgroup of O(2) and is itself  isomorphic to the circle group U(1). Therefore, the SO(2) invariance of the Hamiltonian when expressed with Majorana fermions [Eq.~(\ref{Eqn:Hmajorana})] is equivalent to the U(1) charge conserving symmetry.

Now we show that the discrete reflections, that are part of O(2), correspond to the particle-hole transformation in the fermion language. Let us represent the reflection $\mathbf{P}$ by
\begin{equation}
   \mathbf{P}= \left(
   \begin{array}{c c}
      1 & 0 \\
      0 & -1
   \end{array}
   \right)\;,\quad\mbox{such that}\quad
   \mathbf{P} \left(\!
   \begin{array}{c}
      \gamma\\
      \eta
   \end{array}\!\right)
   = \left(\!
   \begin{array}{c}
      \gamma\\
      -\eta
   \end{array}\!\right)\;.
\end{equation}
From Eq.~(\ref{Eqn:majorana}) one can see that $\mathbf{P}$ swaps creation and annihilation operators. Consequently, the Majorana reflection $\mathbf{P}$ corresponds to the aforementioned particle-hole transformation for fermions on bipartite lattices.

\section{QMC methods} \label{sec:method}

The many-fermion problem defined by the Hamiltonians of Eqs.~(\ref{Eq:HFree})--(\ref{Eq:HV2}) can be solved in many ways, without formally encountering a sign problem. The absence of the sign problem is a necessary, but not a sufficient condition for polynomial scaling of the computational effort. Below we will show that a method of choice such as the continuous-time QMC algorithm in the interaction expansion ($\mbox{CT-INT}$) \cite{Rubtsov05,Gull11} shows fat-tailed distributions for some observables, rendering the central limit theorem inapplicable. To avoid this problem we have used the auxiliary-field QMC method (AF-QMC) \cite{Blankenbecler81,Assaad08_rev} in the Majorana representation. In this section we will first briefly introduce both methods, and then compare them. We finish on a note, suggesting that the continuous-time auxiliary-field algorithm (CT-AUX)\cite{Rombouts99} may be the adequate continuous-time formulation for the problem.  

\subsection{Continuous-time QMC algorithm}

First, we focus on the CT-INT algorithm which stochastically samples the grand-canonical partition function $Z$. The formalism is action based where one distinguishes the Gaussian part $S_0$ and the interaction part $S_I$
\begin{eqnarray}
   S_0 &=& -\sum_{{\bf i},{\bf j}}\iint_0^\beta \dd\tau \, \dd\tau ' c^\dagger_{{\bf i},\tau} G_0^{-1}({\bf i - j}, \tau - \tau')\, c_{{\bf j},\tau'}\quad\quad \\
   S_I &=& V \sum_{\langle {\bf i},{\bf j} \rangle}\int_0^\beta \dd\tau \left(c^\dagger_{{\bf i},\tau}c_{{\bf i},\tau}^{\pd}-\frac{1}{2}\right) \left(c^\dagger_{{\bf j},\tau}c_{{\bf j},\tau}^{\pd}-\frac{1}{2}\right)\, ,
\end{eqnarray}
such that the partition function may be written as the interaction expansion
\begin{equation}
   \label{Eqn:z_ctint}
   Z = \Tr\left[ \E^{-\beta (\mathcal{H}_0+\mathcal{H}_{V})} \right] \\
     = Z_0 \sum_n \frac{\left( -1 \right )^n}{n!}  \left \langle S_{I}{}^n \right \rangle_0 \;,
\end{equation}
where we defined ${\left \langle \dots \right \rangle_0 = Z_0^{-1} \int \mathcal{D}[c^\dagger,c] \left[ T \dots \E^{-S_0} \right]}$ and ${Z_0=\Tr\left[ \E^{-\beta \mathcal{H}_0} \right]}$. Since the expectation value in \Equ{Eqn:z_ctint} is taken with respect to the non-interacting part, we can use Wick's theorem within each term of the Taylor series.  In order to have a sign problem free simulation, the sign of the determinant for the given contribution has to cancel the alternating sign $(-1)^n$ such that the overall value is strictly positive. In \Ref{Huffman14}, Huffman and Chandrasekharan have proven the absence of the sign problem for bipartite lattices with real hopping and repulsive density-density interactions both connecting sites of different sublattices only. Furthermore, the system has to be particle-hole symmetric which restricts the simulations to half-filling. The sign problem free CT-INT simulation is then based on sampling pairs of vertices, such that particle-hole symmetry is ensured for all Monte Carlo configurations. The $t$-$V$ model (\ref{Eq:HFree})--(\ref{Eq:HV2}) fulfills all requirements such that we can employ the CT-INT method.

\subsection{Auxiliary-field QMC algorithm}\label{sec:afqmc}

Secondly, we have implemented an AF-QMC algorithm, similar to Ref.~\onlinecite{Gubernatis85}, but based on the recently presented Majorana QMC method\cite{Li15}. Importantly, this method uses the manifestly O(2) symmetric Hubbard-Stratonovich decomposition, defined on a discretized imaginary-time axis with $\beta=\Delta\tau N_{\tau}$:
\begin{eqnarray}
   \label{eqn_HS}
   \E^{-\Delta\tau V \left(n_{\bf i}-\frac{1}{2})(n_{\bf j}-\frac{1}{2}\right)} 
   & = & \E^{\frac{\Delta\tau V}{2}\left[(c_{\bf i}^{\dagger} c_{\bf j}^{\phantom\dagger} + \text{H.c.})^{2} -\frac{1}{2}\right]  }\\
   & = & \frac{1}{2} \E^{-\frac{V\Delta\tau}{4}}\!\!\sum\limits_{\sigma_{\bf ij}=\pm 1} \E^{-\lambda \sigma_{\bf ij}(c_{\bf i}^{\dagger} c_{\bf j}^{\phantom\dagger} + \text{H.c.}) }\;,\nonumber
\end{eqnarray}
where $\cosh(\lambda)=\exp(V\Delta\tau/2)$. In the following, we partition all nearest-neighbor bonds into ${N_b=4}$ groups: $\mathcal{H}_{0}+\mathcal{H}_{V}=\sum_{b=1}^{N_b}\mathcal{H}_{0}^{(b)}+\mathcal{H}_{V}^{(b)}$. Within each group $\mathcal{M}_{b}$ the bond terms $(c_{\bf i}^{\dagger} c_{\bf j}^{\phantom\dagger} + \text{H.c.})$ commute. 
The  grand-canonical partition function can be written as
\begin{eqnarray}
   Z & = & \text{Tr}\left[ \E^{-\beta (\mathcal{H}_{0}+\mathcal{H}_{V})}\right]\nonumber\\
   & = &\text{Tr}\left[\prod\limits_{l=1}^{N_{\tau}} \E^{-\Delta\tau (\mathcal{H}_{0}+\mathcal{H}_{V})}\right]\nonumber\\
   & = &\text{Tr}\left[\prod\limits_{l=1}^{N_{\tau}} \prod\limits_{b=1}^{N_b}\E^{-\Delta\tau (\mathcal{H}_{0}^{(b)}+\mathcal{H}_{V}^{(b)})}\right]+\mathcal{O}(\Delta\tau^{2})\nonumber\\
   & \propto & \sum\limits_{\{\sigma\}}\text{Tr}\prod\limits_{l=1}^{N_{\tau}}\prod\limits_{b=1}^{N_b} B^{(b)}_{l,l-1}+\mathcal{O}(\Delta\tau^{2})   \label{Eqn:partition_BSS_prefactor}\\
   & = & \sum\limits_{\{\sigma\}} \det\left[\mathds{1}+\prod\limits_{l=1}^{N_{\tau}}\prod\limits_{b=1}^{N_b}{\bf B}^{(b)}_{l,l-1}\right]+\mathcal{O}(\Delta\tau^{2})\;,
   \label{Eqn:partition_BSS}
\end{eqnarray}
where we use
\begin{eqnarray}
   B^{(b)}_{l,l-1}= \prod\limits_{\langle {\bf i j}\rangle\in \mathcal{M}_{b}} \E^{(\Delta\tau t-\lambda \sigma_{{\bf ij}l}) (c_{\bf i}^{\dagger} c_{\bf j}^{\phantom\dagger} + \text{H.c.})}\;,
\end{eqnarray}
and dropped a constant prefactor in Eq.~(\ref{Eqn:partition_BSS_prefactor}) for simplicity. The sum $\{\sigma\}$ extends over the auxiliary-field components $\sigma_{{\bf ij}l}=\pm 1$ on the space-time lattice. We now derive the positivity of the determinants in the partition function in Eq.~(\ref{Eqn:partition_BSS}) of our implementation of the AF-QMC method: The partition function in the Majorana representation reads
\begin{eqnarray}
   \label{Eqn:Z_BSS_2}
   Z 
   &=& \sum\limits_{\{\sigma\}} \text{Tr}\!\!\!\!\prod\limits_{l,b,\langle {\bf ij} \rangle \in \mathcal{M}_{b} }\!\!\!\!
         \E^{a_{{\bf ij} l} (c_{\bf i}^{\dagger} c_{\bf j}^{\phantom\dagger} + \text{H.c.}) } \nonumber\\
   &=& \sum\limits_{\{\sigma\}} \text{Tr}\!\!\!\!\prod\limits_{l,b,\langle {\bf ij} \rangle\in \mathcal{M}_{b} }\!\!\!\!
         \E^{\frac{\I}{2}a_{{\bf ij}l} ( \gamma_{\bf i} \gamma_{\bf j}  + \eta_{\bf i} \eta_{\bf j})  }\nonumber\\
   &=& \sum\limits_{\{\sigma\}} \left[
         \text{Tr}\!\!\!\!\prod\limits_{l,b,\langle {\bf ij} \rangle\in \mathcal{M}_{b} }\!\!\!\!
         \E^{\frac{\I}{2}a_{{\bf ij}l}  \gamma_{\bf i} \gamma_{\bf j}}\right]^{2}\;,
\end{eqnarray}
where  $a_{{\bf{ij}}l}=\Delta\tau t-\lambda \sigma_{{\bf ij}l}$. The absence of negative sign problem amounts to showing that the trace is a real number.  One will readily see this by  reintroducing fermion operators. For ${{\bf i} \in A}$ and ${\bf j} \in B,C$, ${{\frac{\I}{2}a_{{\bf ij}l} \gamma_{\bf i} \gamma_{\bf j}} = \frac{1}{2} a_{{\bf ij}l} ( c^\dagger_{{\bf i}} - c_{{\bf i}}) ( c^\dagger_{{\bf j}} + c_{{\bf j}})}$. Since ${a_{{\bf{ij}}l}\in\mathbb{R}}$, the operator $\E^{\frac{\I}{2}a_{{\bf ij}l}  \gamma_{\bf i} \gamma_{\bf j}}$ is real representable in real space such that the trace will be real. General considerations on  how to avoid the sign problem  within the Majorana representation can be found in Refs.~\onlinecite{Li15,Wang15b,Wei16,Li16}.

\subsection{Comparison of QMC methods}
\begin{figure}[pt]
   \includegraphics[width=0.88\columnwidth]{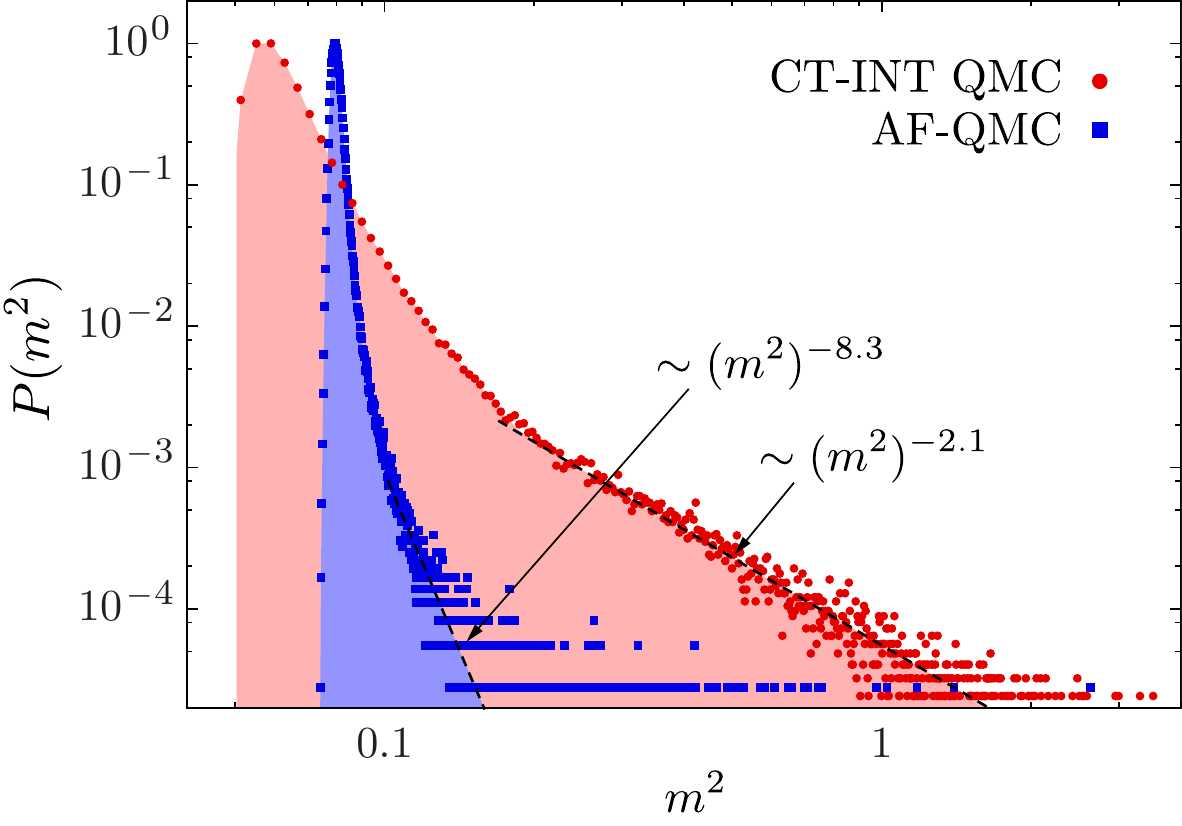}
   \caption{Distribution of the (squared) order parameter for unbinned data from AF-QMC and CT-INT simulations. The data from the CT-INT  simulation shows slowly decaying tails which renders the variance ill-defined.
   \label{Fig:histogramMC}}
\end{figure}
The CT-INT method stochastically evaluates the series expansion for the grand-canonical partition function in Eq.~(\ref{Eqn:z_ctint}) to all orders, and imaginary time can be treated as a continuous parameter. In the AF-QMC algorithm, the path in imaginary time is discretized with a finite resolution $\Delta\tau$, measured in units of inverse energy, $[k_\text{B}T]^{-1}$ [see Eq.~(\ref{Eqn:partition_BSS})]. The systematic error introduced by the discretization is a cutoff at higher and higher energies, as $\Delta\tau\rightarrow 0$. Therefore, it is expected to leave in particular the low-energy physics invariant. For the AF-QMC simulations we use ${\Delta\tau t=0.1}$ throughout the manuscript. 

Interestingly, the different stochastic sampling procedures of the two methods can lead to markedly different distributions of observables. We  observe that the tail of the distribution generated by the $\mbox{CT-INT}$ simulation is decaying considerably slower than it is the case for the AF-QMC simulation (see Fig.~\ref{Fig:histogramMC}). As long as the distribution variance is well defined (less than infinite), the central limit theorem applies. The Monte Carlo time scale necessary to obtain a stable variance however becomes unfavorably large in $\mbox{CT-INT}$ simulations, making the method computationally very expensive for the present model. For certain observables and parameter ranges the CT-INT method produces fat-tailed distributions, which can be avoided by switching to AF-QMC algorithm. While also subject to a skewed distribution of the order parameter, the problem is significantly alleviated. The AF-QMC method may itself be affected by diverging variances which has been studied recently and a remedy has been proposed in Ref.~\onlinecite{Shi16}.

The question then arises how to formulate an efficient sign problem free continuous-time QMC algorithm for the Lieb lattice. Here we briefly argue that the method of choice is the CT-AUX algorithm \cite{Rombouts99,Gull11} which is sign problem free in the Majorana representation. Consider the partition function
\begin{widetext}
\begin{eqnarray}
	 \E^{\beta 4 L^2 K  } Z & = & \text{Tr}\, \E^{ -\beta \left( \mathcal{H}_0+ \sum_{\langle {\bf ij} \rangle}  \left[ V \left(n_{\bf i} - \frac{1}{2}\right) \left(n_{\bf{j}} - \frac{1}{2}\right) - K\right]  \right)  }  \nonumber \\ 
	& = &  Z_0 \sum_{n=0}^{\infty} \frac{(K-V/4)^n}{n!} \int_{0}^{\beta}  \dd\tau_1  \sum_{\langle {\bf i}_1, {\bf j}_1 \rangle} \cdots  \int_{0}^{\beta}  \dd\tau_n  \sum_{\langle {\bf i}_n {\bf j}_n \rangle}  
	\langle T\;\mathcal{H}^{\text{int}}_{\langle {\bf i}_n {\bf j}_n \rangle} \left(\tau_n\right)   \cdots \mathcal{H}^{\text{int}}_{\langle {\bf i}_1 {\bf j}_1 \rangle} \left(\tau_1\right)   \rangle_0 \;,
\end{eqnarray}
\end{widetext}
where $K$ is a real parameter and
\begin{eqnarray}
   \mathcal{H}^{\text{int}}_{\langle {\bf i j} \rangle}
   & = & 1 - \frac{V}{K-V/4} \left[ \left(n_{\bf i} - \frac{1}{2}\right) \left(n_{\bf{j}} - \frac{1}{2}\right)  - \frac{1}{4} \right] \nonumber\\
   & = & 1 + \frac{V}{2K-V/2} 
 \left( c^{\dagger}_{\bf i} c^{}_{\bf j }  + c^{\dagger}_{\bf j} c^{}_{\bf i } \right)^2 \nonumber\\
   & = & \frac{1}{2}\sum_{s =\pm 1} \E^{s  \alpha \left( c^{\dagger}_{\bf i} c^{}_{\bf j }  + c^{\dagger}_{\bf j} c^{}_{\bf i } \right)} \;.
\end{eqnarray}
The last identity follows from the special form of the interaction which satisfies 
\begin{equation}
   \left( c^{\dagger}_{\bf i} c^{}_{\bf j }  + c^{\dagger}_{\bf j} c^{}_{\bf i } \right)^4 = \left( c^{\dagger}_{\bf i} c^{}_{\bf j }  + c^{\dagger}_{\bf j} c^{}_{\bf i } \right)^2 \;,
\end{equation}
and hence the necessary choice
\begin{equation}
   \frac{V}{2K-V/2} +  1 = \cosh(\alpha). 
\end{equation}
Thereby the algorithm can only be formulated  for ${K > V/4}$. 
In general, the parameter $K$ can be tuned to maximize the efficiency of the algorithm and to avoid numerical instabilities due to nearly singular matrices \cite{Gull11}.
With this formulation, the partition function reads
\begin{widetext}
\begin{equation}
   \E^{\beta 4 L^2 K  } Z = Z_0 \sum_{n=0}^{\infty} \frac{(K-V/4)^n}{2^n\, n!} \int_{0}^{\beta}  \dd\tau_1  \sum_{ \langle {\bf i}_1, {\bf j}_1 \rangle, s_1 } \!\!\!\cdots  \int_{0}^{\beta}  \dd\tau_n  \sum_{\langle {\bf i}_n, {\bf j}_n \rangle, s_n}  
   \langle T\;  \E^{s_n \alpha \left( c^{\dagger}_{{\bf i}_n} c^{\phantom\dagger}_{{\bf j }_n}  + c^{\dagger}_{{\bf j}_n} c^{\phantom\dagger}_{{\bf i }_n} \right)(\tau_n)} \cdots  
   \E^{s_1 \alpha \left( c^{\dagger}_{{\bf i}_1} c^{\phantom\dagger}_{{\bf j }_1}  + c^{\dagger}_{{\bf j}_n} c^{\phantom\dagger}_{{\bf i }_1} \right)(\tau_1)} \rangle_0.
\end{equation}
\end{widetext}
The absence of the sign problem again follows directly from the Majorana representation discussed in Sec.~\ref{sec:afqmc}. Using arguments presented in Ref.~\onlinecite{Mikelsons09}, the relation between the CT-AUX formulation and the AF-QMC algorithm becomes apparent in the respective limits ${K \rightarrow \infty}$ and  ${\Delta \tau \rightarrow 0}$. In these limits the number of vertices in the CT-AUX algorithm [which scales as $\mathcal{O}(K)$ (see \Ref{Assaad07})] diverges and they are homogeneously distributed in the imaginary-time interval ${\left[0, \beta \right]}$. Such a distribution is also achieved in the limit ${\Delta \tau \rightarrow 0}$. It is also interesting to note that the constraint ${K > V/4}$ implies that there is no  path interpolating from the CT-INT (${K=0}$) to the CT-AUX algorithm. 

Generically, the continuous-time methods scale as the Euclidean volume cubed. Recently alternative formulations of the sampling \cite{Wang15,Iazzi15} allow for a linear scaling in inverse temperature and thereby place the  continuous-time methods formally in the same efficiency class as the auxiliary-field method albeit without the systematic Trotter discretization error. Since the linear in $\beta$ CT-INT and CT-AUX approaches sample the very same configuration space as the generic continuous-time methods, we expect the same distributions to occur in both formulations.

\section{Results} \label{sec:results}

\begin{figure}[pt]
   \includegraphics[width=0.9\columnwidth]{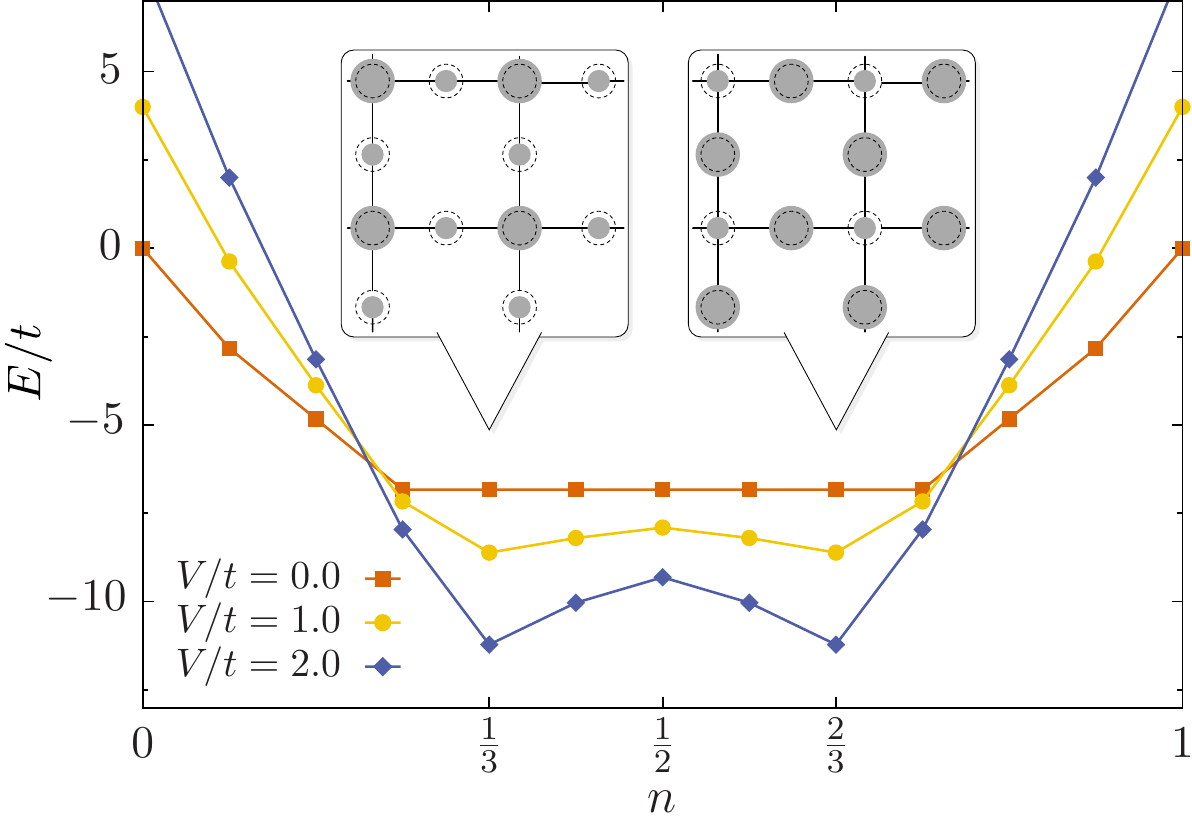}
   \caption{Ground-state energy $E$ for fixed electron density $n$ and various interaction strength $V$ on a ${L=2}$ lattice from exact diagonalization. The distribution of the charge density for the ground state with $1/3$- and $2/3$-filling is illustrated in the insets. Lines are guide to the eye only.
   \label{Fig:EDResults}}
\end{figure}

Due to the symmetry protected flat band at the Fermi level we expect a high susceptibility to interaction effects.  We employ the exact diagonalization method to get a first insight. The ground-state energy versus the filling for various interaction strengths $V$ on a lattice with linear dimension ${L=2}$ is shown in \Fig{Fig:EDResults}. We observe the expected thermodynamic instability driving the system away from the particle-hole symmetric point at half-filling towards a filling fraction of either $1/3$ or $2/3$. The ground state is unique for the two filling fractions stated above. For any other finite amount of particles the ground state is multiply degenerate. The insets depict the density distribution of ground state wave function for the $N=4$ and $N=8$ sector which is proportional to the area of the gray discs. It illustrates the formation of the charge order which accumulates the electrons on the sublattice $A$ ($B$ and $C$) for a filling fraction of $1/3$ ($2/3$). The energy of both states is degenerate, such that particle-hole symmetry may be broken spontaneously in the thermodynamic limit.

\begin{figure}[pt]
   \includegraphics[width=\columnwidth]{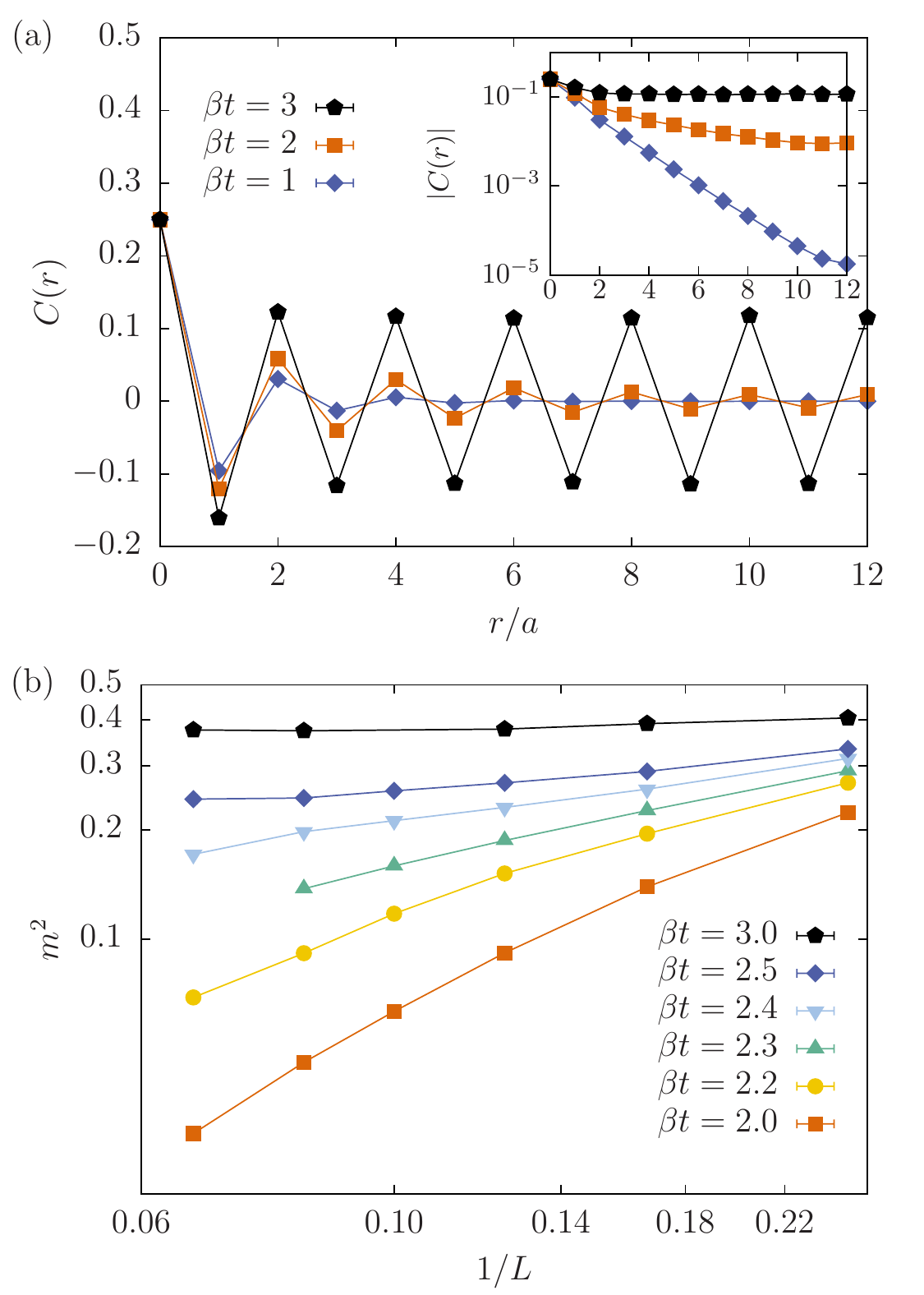}
   \caption{Density correlation function for ${L=12}$ and at ${V/t=2}$ as a function of spatial separation along the lattice axes (a). 
The inset displays the growing correlation length on  a semi-logarithmic scale. 
Panel (b) shows the the finite-size behavior of the (squared) order parameter on a double-logarithmic scale. 
The data is compatible with a phase transition between $\beta t=2.4$ and $\beta t = 2.5$.\label{Fig:CDWcorr}}
\end{figure}

In the following we will confirm this intuition by studying larger lattice sizes with the QMC methods presented in Sec.~\ref{sec:method}. We therefore begin with the analysis of charge correlation functions and continue with the extraction of the corresponding order parameter and a finite-size scaling to investigate the critical behavior. Due to the severe tail of the distribution in \Fig{Fig:histogramMC} we used the AF-QMC method unless it is stated otherwise. We measure the density correlations
\begin{equation}
   C(r) =\frac{1}{4 L^{2}}\sum\limits_{\mathbf{i},\mathbf{j}}
   \left(\langle n_{\mathbf{i}}n_{\mathbf{j}}\rangle - \langle n_{\mathbf{i}}\rangle\langle n_{\mathbf{j}}\rangle\right) \delta(|\mathbf{i}-\mathbf{j}|-r)\;,
\end{equation}
where $\mathbf{i}-\mathbf{j}$ is taken along  the lattice axes $\mathbf{a}_{1,2}$. Figure~\ref{Fig:CDWcorr}(a) shows the spatial pattern of the charge distribution and the inset shows the decay of $|C(r)|$ on a semi-logarithmic scale from which a growing correlation length for lower temperatures can be inferred. This clearly confirms the formation of the expected charge order at low temperatures ${\beta=3}$ (${T=0.33}$). 
Additionally, we have also measured current correlations to test for the emergence of a 
quantum anomalous Hall state. The absence of a signal in the current correlations allows 
us to exclude this competing order and hence the occurrence of time-reversal symmetry 
breaking (not shown).

\begin{figure*}[pt]
   \includegraphics[width=2.0\columnwidth]{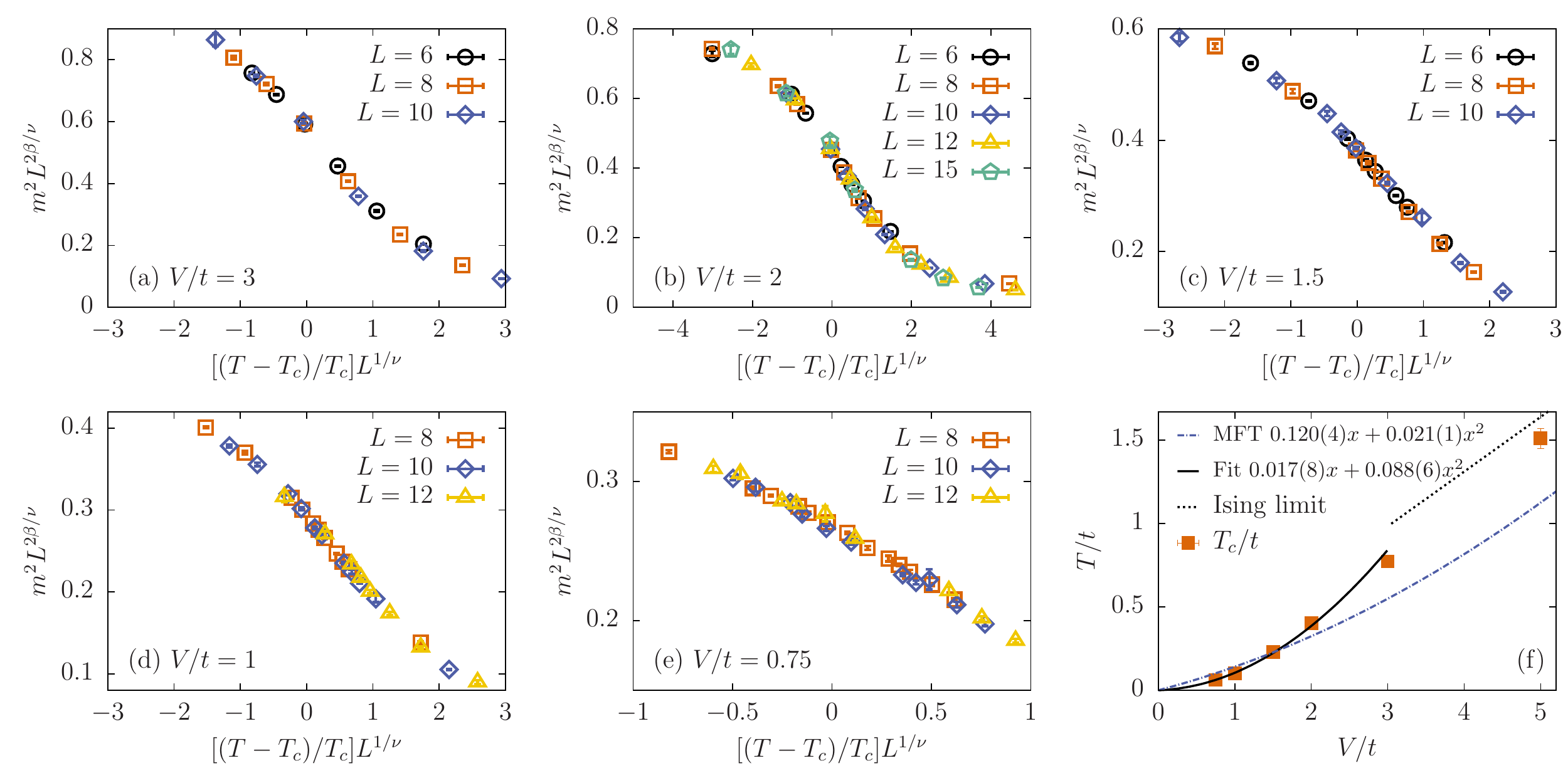}
   \caption{Finite-size data collapses (a--e) of the (squared) order parameter across the thermal phase transition using the critical exponents of the two-dimensional Ising model for different interaction strength $V$. 
   Panel (f) shows the behavior of the extracted critical temperatures as a function of $V$ compared to mean-field theory (MFT) results 
   and the classical Ising-limit for strong Coulomb repulsion \cite{footnote_fig5}.
   \label{Fig:DataCollapse}}
\end{figure*}
\textit{Finite-Temperature phase transition} -- We measured the order parameter $m$ to study the phase transition between the unstable metal and the charge-ordered phase:
\begin{equation}
   m(T,L)=\sqrt{\text{Tr}[\mathbf{N}(\mathbf{Q})]}\;,
\end{equation}
where $\mathbf{Q}=(0,0)$ and the density-density correlation function $\mathbf{N}(\mathbf{Q})$ is a $3\times 3$-matrix. Its elements are
\begin{equation}
   N_{ab}(\mathbf{Q})=\frac{1}{L^2}\sum\limits_{x}^{L^2} \langle n_{1}^{a} n_{x}^{b}\rangle -\langle n_{1}^{a}\rangle \langle n_{x}^{b}\rangle\;,
\end{equation}
where $n_{x}^{a}$ is the fermion density of orbital $a=\{1,2,3\}$ in unit cell $x=\{1,\cdots,L^{2}\}$. We have studied lattices of linear length $L=4,6,8,10,12,\text{ and }15$. The finite-size behavior of the (squared) order parameter, shown in  \Fig{Fig:CDWcorr}(b) for $V/t=2$, suggests that it acquires a finite value for inverse temperatures above $\beta t=2.4$. We test the hypothesis that the phase transition belongs to the two-dimensional Ising universality class by performing a finite-size scaling analysis for $N_{\text{data}}$ sets of measurements $\{m_{i},T_{i},L_{i}\}$, at a given interaction strength~$V$. The Ansatz is
\begin{equation}
   \label{eqn_fss_ansatz}
   m_{i}(T_{i},L_{i})=L_{i}^{-\beta/\nu}\mathcal{F}[(T_{i}-T_{c})/T_{c}\;L_{i}^{1/\nu}]\;,
\end{equation}
where $\mathcal{F}$ is an unknown scaling function. We fix the critical exponents $\beta$ and $\nu$ to their two-dimensional Ising values, $\beta=1/8$ and $\nu=1$, and extract from Eq.~(\ref{eqn_fss_ansatz}) the value of $T_{c}$ which gives the best data collapse for various lattices sizes and temperatures close to the estimated critical temperature. We obtain the estimate $T_{c,\text{est}}$ from the crossing point of $mL^{\beta/\nu}(T)$ using the largest and the second-largest lattice size. We then choose a data window $[T_{\text{min}},T_{\text{max}}]$, with $T_{\text{min}}<T_{c,\text{est}}<T_{\text{max}}$, and fit the data $m_{i}(T_{i},L_{i})L_{i}^{\beta/\nu}$ to a low-order ($k_{\text{max}}\leq 4$) polynomial of the form:
\begin{equation}
   \tilde{\mathcal{F}}(x)=\sum\limits_{k=0}^{k_{\text{max}}}u_{k}\,x^{k}\;.
\end{equation}
To measure the goodness of fit, we compute the statistic $\chi^{2}/\mbox{d.o.f.}$, using 
\begin{equation}
   \chi^{2}=\sum\limits_{i=1}^{N_{\text{data}}}\left\{\frac{m_{i}(T_{i},L_{i})L_{i}^{\beta/\nu}-\tilde{\mathcal{F}}[(T_{i}-T_{c})/T_{c}\;L_{i}^{1/\nu}]}{\sigma_{m,i}}\right\}^{2}\;,
\end{equation}
where $\sigma_{m}$ is the statistical error of the Monte Carlo measurement of the order parameter $m$. We  have also compared our finite-size scaling method to a recently proposed method \cite{Harada11}, based on Bayesian statistics, 
and we obtained the same critical temperatures within error bars (not shown).

We list the fit parameters in Tab.~{\ref{tab:fit}} and  present the resulting data collapse of the squared order parameter in \Fig{Fig:DataCollapse} for various interaction strengths. First this nicely confirms the compatibility with the two-dimensional Ising universality class for this phase transition. Secondly, we can extract the critical temperature $T_c(V)$ for the given interaction strength $V$ which is summarized by \Fig{Fig:DataCollapse}(f) \cite{footnote_fig5}. 
According to the Stoner criterion for the weak coupling regime, one would expect a linear relation ${V \sim T}$ resulting from  the Curie-type non-interacting susceptibility for localized states. This behavior can indeed be seen in the standard charge-order mean-field approximation. 
However, as a consequence of many-body correlations, the QMC simulation result has a dominating quadratic contribution. For strong Coulomb repulsion ${|V| \gg 1}$ the mean field Ansatz can be expected to again correctly describe a linear behavior of $T_c$. Furthermore, in this limit the $t$-$V$ model maps onto a classical two dimensional Ising model, such that the  critical temperature ${T_{c,t\mbox{-}V} = T_{c,\text{Ising}} |V|/4}$ \cite{Gubernatis85}. For the Ising model on the Lieb lattice we have computed the estimate ${T_{c,\text{Ising}} = 1.310(1)}$. The slope of the Ising-limit is indicated in \Fig{Fig:DataCollapse}(f).

\begin{table}[h]
   \begin{tabular}{c|c|c|c|c|c}\hline
     $V/t$ & 0.75 & 1 & 1.5 & 2 & 3 \\\hline \hline
     $T_{c}/t$ & 0.0619(2) & 0.1029(2) & 0.2277(5) & 0.4013(6) & 0.773(2) \\
     $k$ & 3 & 4 & 3 & 3 & 2 \\ 
     $\chi/\mbox{d.o.f.}$ & 21.92/15 & 41.21/20 & 18.24/12 & 19.35/14 & 4.45/8 \\\hline
   \end{tabular}
   \caption{For each interaction strength $V$ the critical temperature $T_{c}/t$ is obtained from the fit of the (squared) order parameter to a polynomial of order $k$.
   \label{tab:fit}}
\end{table}

\begin{figure}[pt]
   \includegraphics[width=\columnwidth]{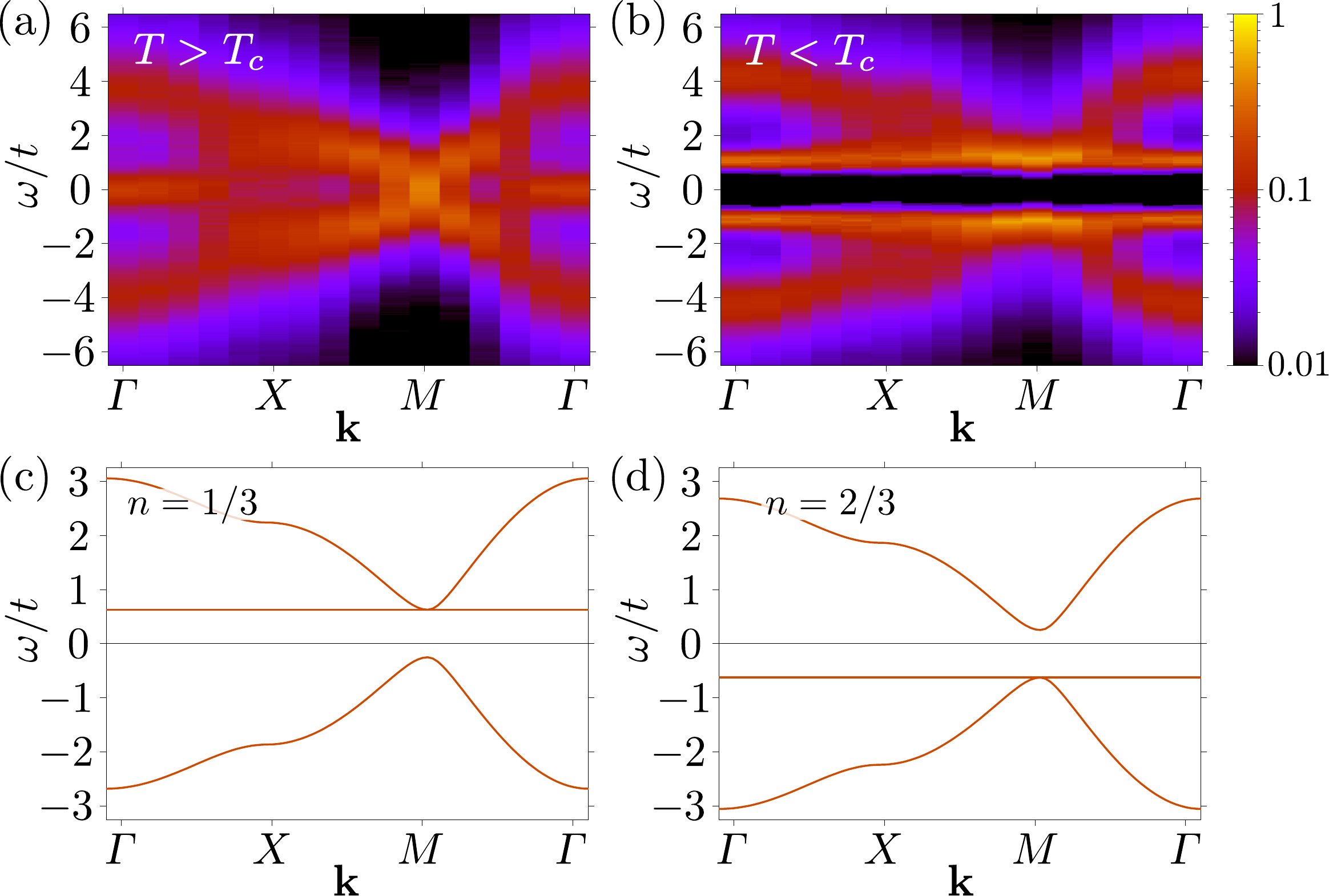}
\caption{The one-particle spectral function at temperature (a) $T=0.5$ above $T_c$ and (b) $T=0.2$ below the Ising transition for a ${L=10}$ lattice and $V/t=2$. The spectrum in (b) can be interpreted as the superposition of the dispersions for (c) 1/3-filling and (d) 2/3 filling (here obtained from mean-field calculations).
   \label{Fig:SpecAboveTc}}
\end{figure}

\textit{Dynamics} -- The phase transition is equally observed in the single-particle excitations. We have obtained $A({\bf k},\omega)$ by analytic continuation of the thermal imaginary-time displaced Green function $G({\bf k},\tau)$, using the stochastic maximum entropy method \cite{Sandvik98,Beach04}. The dynamic Green functions have been measured with the CT-INT method. Above the critical temperature, the single-particle spectrum features three quasiparticle bands [Fig.~\ref{Fig:SpecAboveTc}(a)]. Their form is essentially given by the non-interacting dispersion relation (cf. Fig.~\ref{Fig:FreeSystem}). The overall band width is renormalized with respect to the free system and the single-particle excitations are broadened by temperature and finite lifetime. Importantly, the flat band at ${\omega=0}$ survives in the high-temperature and strongly interacting phase. Below the critical temperature, the model acquires a single-particle gap [Fig.~\ref{Fig:SpecAboveTc}(b)], corresponding to the insulating charge-ordered state.

The breaking of particle-hole symmetry is accompanied by a simultaneously breaking of the balance between electron and hole densities (see Fig.~\ref{Fig:EDResults}), resulting in two orthogonal ground states with electronic densities $n=1/3$ and $n=2/3$. Since we restrict the QMC simulation to half-filling ($n=1/2$), none of the two ground states can be accessed individually. Instead, the QMC algorithm always computes a particle-hole symmetric imaginary-time displaced Green function $G_{n=1/2}({\bf k},\tau)$. Below the critical temperature, we can interpret the Green function as 
\begin{equation}
   G_{n=1/2}({\bf k},\tau)=\frac{1}{2}\left[G_{n=1/3}({\bf k},\tau)+G_{n=2/3}({\bf k},\tau)\right]\;.
\end{equation}
The spectrum $A({\bf k},\omega)$ in Fig.~\ref{Fig:SpecAboveTc}(b) may hence be seen as the equal superposition of the two spectra of the two commensurable charge-ordered states, which are connected by particle-hole transformation. Corresponding dispersions, as obtained from the self consistent mean-field solution for the charge-ordered states, are shown in Fig.~\ref{Fig:SpecAboveTc}(c-d). 

\section{Conclusion} \label{sec:conclusion}

The $t$-$V$ model of spinless fermions on the Lieb lattice provides a technical challenge for QMC simulations but also empowers the very same to reveal properties of phase space previously thought to be unaccessible. In this manuscript we presented the troublesome distribution of observables in CT-INT simulations and compared them against the more favorable AF-QMC algorithm. Furthermore we conjectured the CT-AUX algorithm to be the more efficient implementation in the arsenal of continuous-time QMC algorithms. By means of ED we provided evidence that the half-filled ground state is inherently unstable towards commensurable charge order, which is incompatible with half-filling on the Lieb lattice. While the simulations were constrained to half-filling we have provided strong evidence for a spontaneously broken particle-hole symmetry in the thermodynamic limit by studying the squared order parameter. We also studied the finite-temperature phase transition from thermal disorder to the charge-ordered state at low temperature and determined the critical theory to be compatible with the two-dimensional Ising universality class. The behavior of the critical temperature at weak couplings shows a dominant ${T_c^{\text{QMC}} \propto V^2}$ scaling which stands at odds with the mean-field expectation $T_c^{\text{MF}} \propto V$. Furthermore our simulations allowed us to extract spectral properties away from half-filling: The single-particle spectra may be interpreted as the equal superposition of dispersions at different electronic densities related by a particle-hole transformation. We conjecture that the physics of the $t$-$V$ model on the Lieb lattice extends to other bipartite lattices with a similar, dispersion-less band at the particle-hole symmetric point, such as the Dice lattice.

The unique aspect of our model is that the symmetry broken state  generates a non-zero expectation value of ${\langle n_i - 3/2 \rangle}$, where $n_i$ is the total charge per unit cell. The question arises if it is possible to use this aspect of the Lieb lattice to spontaneously generate chemical potential terms. In particular one can conceive a bilayer system where the first layer is described by the $t$-$V$ model on the Lieb lattice and the second layer is the model of interest which is  assumed to have an SU($N$) symmetry. An interlayer coupling of the form ${\mu (n_{i}^{(1)} - 3/2) (n_{i}^{(2)} - N/2)}$ is particle-hole symmetric. Below $T_c$ -- provided that the interlayer coupling does not alter the nature of the Ising transition --  spontaneous charge ordering will imprint a chemical potential term on the second layer.  \footnote{Alternatively one can break particle-hole symmetry explicitly on the Lieb lattice, by adding a staggered chemical potential. Within the Majorana  representation, this will not introduce a sign problem (cf. Ref.~\onlinecite{Wei16}).} This provides an intriguing possibility to access finite doping by simulating a larger system at half-filling.

\begin{acknowledgments}

We thank S. Chandrasekharan and F. Parisen Toldin for useful discussions.  MB thanks the  Bavarian Competence Network for Technical and Scientific High Performance Computing (KONWIHR) for financial support. JSH thanks the DFG-funded SFB-1170 for  financial support.  FFA and TCL thank the DFG-funded FOR1807 for partial financial support. The authors gratefully acknowledge the computing time granted by the John von Neumann Institute for Computing (NIC) and provided on the supercomputer JURECA \cite{Jureca16} at J\"ulich Supercomputing Centre (JSC). The authors gratefully acknowledge the Gauss Centre for Supercomputing e.V. (www.gauss-centre.eu) for funding this project by providing computing time on the GCS Supercomputer SuperMUC at Leibniz Supercomputing Centre (LRZ, www.lrz.de).
\end{acknowledgments}

\bibliographystyle{apsrev4-1}
\bibliography{paper}

\begin{thebibliography}{55}
\bibitem [{foo()}]{footnote_fig5}%
  \BibitemOpen
  \href@noop {} {}\bibinfo {note} {At $V/t=5$, the resolution $\Delta\tau
  t=0.1$ of inverse temperatures is too coarse for an accurate fit of the
  scaling function. Therefore we used the crossing points of
  $mL^{\beta/\nu}(T)$ together with a conservative error to estimate
  ${T_{c}(V/t=5)=1.51 \pm 0.06}$.}\BibitemShut {Stop}%
\end{thebibliography}

\end{document}